\documentclass[twocolumn,showpacs,amsmath,amstex,amssymb,prb]{revtex4}
\usepackage{graphicx,bm,units}

\newcommand{\brr}{{\bm r}}

\begin{document}

\title{Tunneling transport in devices with semi-conducting leads}

\author{Emil Prodan and Amy LeVee}
\address{Department of Physics, Yeshiva University, New York, NY 10016} 

\begin{abstract}
 The present paper extends the modern theory of tunneling transport to finite temperatures and presents an application for molecular devices made of alkyl chains connected to silicon nano-wires, mapping their transport characteristics as functions of temperature and alkyl chain's length. Based on these calculations and on the analytic theory, it is found that the tunneling decay constant is determined not by the Fermi level, but by the edge of the valence or conductance band, whichever is closer to the Fermi level. Further insight is provided by mapping the evanescent transport channels of the alkyl chains and few other physical quantities appearing in the analytic formula for conductance. A good qualitative agreement with the experimental data is obtained.
 
\end{abstract}


\date{\today}

\maketitle

Our understanding of charge transport through single molecules is growing rapidly due to advances on both experimental and theoretical fronts. This is especially true for molecular devices connected to metallic leads. However, the theoretical understanding of transport in devices connected to semi-conducting leads is happening at a much slower pace. The number of atomistic simulations for this class of devices is quite limited.\cite{Rakshit:2004sy, Meunier:2005dz, Lu:2005yb, Quek:2007lo, Kirczenow:2005ph,Kirczenow:2009cy} This is somewhat surprising giving the fact that today's electronic devices are mostly made of semi-condicting materials. It is also surprising because the self-assembled alkyl chains on silicon surfaces were among the first molecular electronic devices to display highly reproducible transport and characterization measurements.

The past few years have seen the development of what we call the modern theory of tunneling transport.\cite{Mavropoulos:2000cr,Tomfohr:2002oq,Tomfohr:2004ve,Fagas:2004uq} This theory connects the tunneling exponent $\beta$ to the complex band structure of the insulating chains, an approach that goes well beyond the effective mass treatments. In Refs.~\onlinecite{Prodan:2007qv} and \onlinecite{Prodan:2009cs}, one of the authors contributed to the theory by deriving an explicit expression for the contact conductance.  The theory has been applied extensively to devices with metallic leads, but never to devices with semi-conducting leads. For the later case, the following questions are open: a) At what energy should be $\beta$ evaluated? b) Which states determine the contact conductance? c) How does the temperature enter in all of these? The goal of this paper is to clarified all these issues and to show the theory at work for the electronic devices discussed below.  

Electronic devices made of alkyl chains self-assembled directly on silicon surfaces present several key features allowing accurate and reproducible transport measurements and characterizations.\cite{Salomon:2003gd,A.-Salomon:2005kx,Nesher:2006ys,Seitz:2006rw,Seitz:2007dk,Salomon:2007qr,Thieblemont:2008hl,Aswal:2006rm} For example, silicon presents a smooth, reproducible and well-controlled solid surface.\cite{Wade:1997le,Allongue:2000xd} Also, the well-prepared Si-C bonded alkyl monolayers are stable,\cite{Sieval:2000kx} and suitably long alkyl chains form well-organized, nearly pinhole-free monolayers.\cite{Sieval:2000kx,Sieval:1999ys,Sieval:1998zr} The highly covalent Si-C bond, unlike the metal-thiolate or metal-amine bond etc., is a natural continuation of the alkyl chain, a feature that minimizes potential discontinuities at the contacts.\cite{Thieblemont:2008hl,Salomon:2007qr,Salomon:2003gd} The absence of a linker between Si and the alkyl chain allows the direct study of the electrical properties of the monolayer itself. The systems themselves are interesting because the choice of Si electrodes makes the devices relevant to Si-based microelectronics.

The electronic structure of self-assembled alkyl chains on Si(111) surface was studied theoretically in Ref.~\onlinecite{Segev:2006vn}. No such study exists for the charge transport properties of these devices. Using geometries taken from experiments and previous theoretical works,\cite{Sieval:2000vn,Sieval:2001uq} Ref.~\onlinecite{Segev:2006vn} provides a good correlation between the theoretically computed and experimentally measured ultraviolet photoelectron spectra (UPS) and inverse photoemission spectroscopy (IPES). To obtain simultaneous agreement with the two sets of data, the theoretical Si insulating gap had to be properly adjusted by a rigid shift of the conduction states and the valence states had to be slightly stretched. These corrections were mainly necessary to correct for the smaller theoretical Si gap, a well known shortcoming of the current Density Functional Theory (DFT) kernels. However, the electronic structure around the band edges seemed to be reasonably well reproduced by these calculations, leading us to believe that DFT is appropriate for studying the transport when the  Fermi level is close to the valence or conduction bands. Another interesting output of the calculations of Ref.~\onlinecite{Segev:2006vn} was that the lateral dispersion of the energy bands of the alkyls was  weak, implying that the electronic states are localized on the chains even at relatively high monolayers coverage. This suggests that the transport properties of the monoloyers can be analyzed using just a single chain.

The experiments mentioned above involve long alkyl chains, typically containing more than 12 methyl groups. We compute the transport characteristics of devices with only up to 8 methyl groups, but we show that the systems already reached the asymptotic regime and therefore we can extrapolate the conductance to larger devices.  To compute the tunneling conductance of such large molecular devices, we use the asymptotic expression of the tunneling conductance derived in Refs.~\onlinecite{Prodan:2007qv} and \onlinecite{Prodan:2009cs}. We extend these derivations to finite temperature and then apply the new formalism to study the tunneling properties of single alkyl chains connected to semi-conducting Si nano-wires. The Si nano-wires were chosen because of certain computational advantages. Due to the lateral confinement of the electrons in Si nano-wires, the electronic structure of the electrodes quantitatively and qualitatively differs from that of bulk Si. Nevertheless, the present simulations allow us to map and rationalize the dependence of the transport characteristics on temperature and alkyl chain's length and to make qualitative comparisons with the experiments.

\begin{figure}
\center
\includegraphics[width=8.6cm]{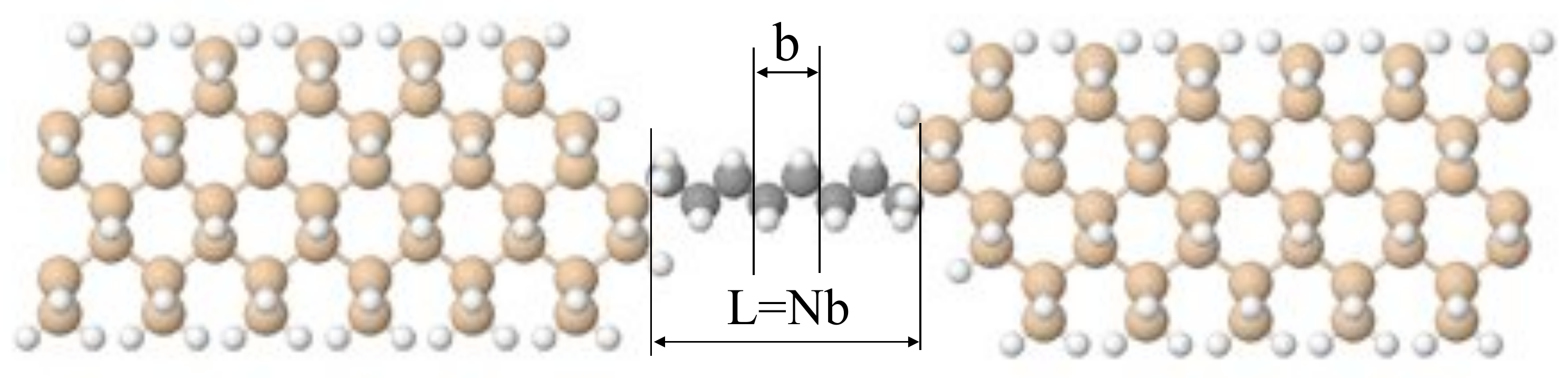}
\caption{(Color online) Illustration of a typical device considered in this paper. It contains an alkyl chain covalently bonded to Si nano-wire leads. The figure indicates the unit cell that is repeated to obtain the periodic potential $V_0$. It also defines the length $L$ of the chain.}
 \label{setup}
\end{figure}

\section{Theoretical Framework}

We consider a charge transport experiment involving a device made of an insulating molecular chain attached to semi-conducting leads (see Fig.~\ref{setup}). The distinction between insulating and semi-conducting chains is that the former have much larger energy gaps. The device is assumed oriented along the $z$ axis. The charge current is driven by a small time oscillating electric field ${\bf E}^{\mbox{\tiny{ext}}}({\bf r})e^{i\omega t}$, whose effects are treated in the linear response regime. The dc regime is obtained by letting the frequency of the oscillations go to zero. The existence of a steady state is implicitly assumed.

The device will not conduct electricity at zero temperature. However, we can simplify the discussion if we start from the zero temperature formalism and assume metallic leads. For this case it was recently established that,\cite{Prodan:2009cs} within the Time Dependent Current-Density Functional Theory,\cite{G.-Vignale:1996fk,Vignale:1997fk} the linear conductance of a molecular device is given by: 
 \begin{equation}\label{exactg}
 G = \int d \brr_\bot   \int d \brr'_\bot  \  [(1- \hat{\sigma}^{\mbox{\tiny{KS}}}*\hat{{\cal F}})^{-1}* \hat{\sigma}^{\mbox{\tiny{KS}}}]_{zz}(\brr,\brr';\epsilon_{\text{\tiny{F}}}).
 \end{equation}
 Here, $\hat{{\cal F}}$ is the matrix kernel
 \begin{equation}
{\cal F}_{\alpha \beta}(\brr,\brr') \equiv  \left . \frac{\delta E_\alpha^{\mbox{\tiny{dyn}}}(\brr)}{\delta j_\beta (\brr')}\right |_{{\bf j}=0},
\end{equation}
$E_\alpha^{\mbox{\tiny{dyn}}}(\brr)$ being the dynamical response of the electrons to the external perturbation,\cite{G.-Vignale:1996fk,Vignale:1997fk} and $\hat{\sigma}^{\mbox{\tiny{KS}}}_{zz}(\brr,\brr';\epsilon_{\text{\tiny{F}}})$ is the local conductivity tensor of the non-interacting Kohn-Sham electrons. $\brr_\bot$ and $\brr'_\bot$ denote the coordinates of two arbitrary planes normal to the axis of the device.

In the tunneling experiments, the currents are extremely small and the non-adiabatic effects are expected to be negligible. In this case, we can neglect the term $\hat{\sigma}^{\mbox{\tiny{KS}}}$$*$$\hat{{\cal F}}$ and the zero temperature linear conductance takes the simple form
 \begin{equation}\label{exactg}
 G = \int d \brr_\bot   \int d \brr'_\bot  \   \hat{\sigma}^{\mbox{\tiny{KS}}}_{zz}(\brr,\brr';\epsilon_{\text{\tiny{F}}}).
 \end{equation}
This expression was previously derived in Refs.~\onlinecite{Fisher:1981kx} and \onlinecite{Baranger:1989bs} for non-interacting electrons  and it is equivalent to the Landauer formula applied to the non-interacting Kohn-Sham electrons. Therefore, Refs.~\onlinecite{Prodan:2007qv} and \onlinecite{Prodan:2009cs} gives a formal justification why Landauer formalism is appropriate for treating the off-resonance tunneling transport of interacting electrons. 

Refs.~\onlinecite{Fisher:1981kx} and \onlinecite{Baranger:1989bs} also established the finite temperature expression for the conductance of non-interacting electrons. We arrived at the same expression after including  the finite temperature in the derivations of Refs.~\onlinecite{Prodan:2007qv} and \onlinecite{Prodan:2009cs} that lead to Eq.~\ref{exactg}. This expression is:
\begin{equation}\label{cond}
G=\frac{2e^2}{h}\int d\epsilon [-f'_{\text{\tiny{FD}}}(\epsilon-\epsilon_F)] \ T(\epsilon)
\end{equation}
with
\begin{equation}\label{trans}
T(\epsilon) \equiv \left ( \frac{2e^2}{h} \right ) ^{-1}\int d \brr_\bot  \int d\brr'_\bot \ \sigma_{zz}^{\mbox{\tiny{KS}}}(\brr,\brr';\epsilon)
\end{equation}
Here, $f'_{\text{\tiny{FD}}}(\epsilon)$ is the derivative of the Fermi-Dirac distribution: $f_{\text{\tiny{FD}}}(\epsilon)=(1+e^{\beta \epsilon})^{-1}$ [$1/\beta$=Boltzmann constant $\times$ Temperature]. The above expression applies equally well to devices with metallic or semi-conducting leads. To avoid confusion, we use $\tau$ to denote the temperature.

\begin{figure}
  \includegraphics[width=8.6cm]{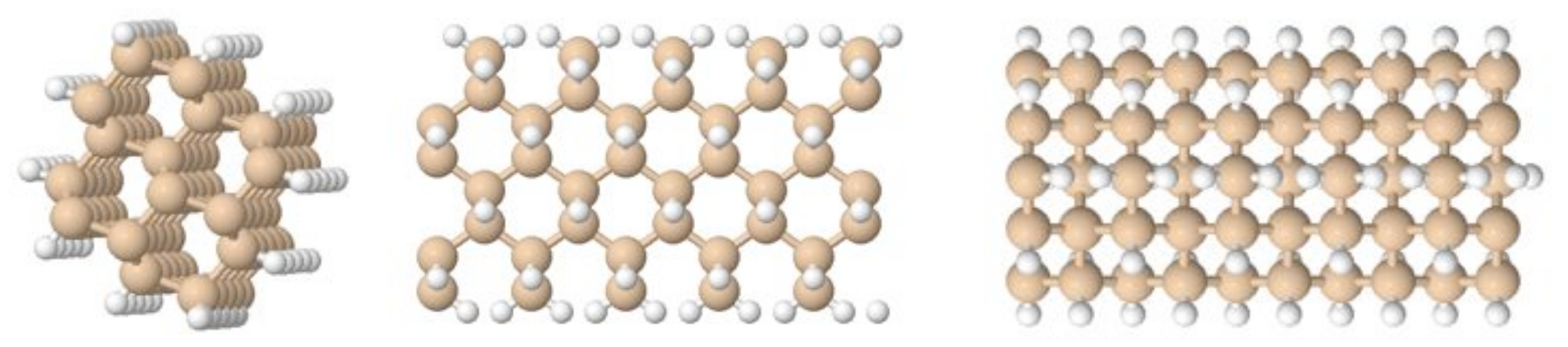}
  \caption{(Color online) The atomic configuration of the Si wires. The left diagram shows a view along the wire and the next two show views from the sides.}
 \label{siwire}
\end{figure}

The exact asymptotic expression, in the limit of long alkyl chains, of $T(\epsilon)$ was computed in Refs.~\onlinecite{Prodan:2007qv} and \onlinecite{Prodan:2009cs}. Let us give a brief discussion of this result. Assume that a self-consistent Kohn-Sham calculation has been completed for the entire device. As we shall see later in this paper, to a high degree of accuracy, the effective potential $V_{\mbox{\tiny{eff}}}$ of the entire molecular device  can be decomposed into a perfectly periodic piece $V_0$, extending from negative to positive infinity, and a difference $\Delta V$=$V_{\mbox{\tiny{eff}}}-V_0$.  The periodic potential $V_0$ is constructed by periodically repeating the effective potential between $-b/2$ and $b/2$ in the middle of the chain, as indicated in Fig.~\ref{setup}. It is useful to regard the self-consistent Kohn-Sham Hamiltonian of the entire device as a periodic Hamiltonian,
 \begin{equation}\label{hzero}
 H_0=-\frac{\hbar^2}{2m}\nabla^2+V_0(\brr), \ V_0(\brr + b{\bf e}_z)=V_0(\brr),
 \end{equation}
strongly perturbed by the leads via the potential $\Delta V$, which is localized on the electrodes and is practically zero inside the alkyl chain. The effective Hamiltonian of the entire system is then
 \begin{equation}\label{hamil}
 H=H_0+\Delta V_{\mbox{\tiny{L}}}(\brr)+\Delta V_{\mbox{\tiny{R}}}(\brr),
 \end{equation}
where we divided  $\Delta V$ into the left and right parts relative to the mid point of the device.

The Bloch functions for the periodic potential $V_0$ are usually characterized by a $k$-vector and a band index. Since here we consider energies inside the insulating gap of $V_0$, the band index losses its precise definition and it is more useful to drop this index and see the $k$-vector of the Bloch functions as defined on a Riemann surface. This Riemann surface was described extensively in Ref.~\onlinecite{Prodan:2006yq}. It turns out that the electron band dispersion $\epsilon_k$ can be extended to the whole Riemann surface and if we pick an arbitrary energy (possibly complex like $\epsilon_{\mbox{\tiny{F}}}+i\delta$) then the equation $\epsilon_k=\epsilon$ has a sequence $\{k_\alpha\}$ of infinite solutions. We are interested in the case when $\epsilon = \epsilon_{\mbox{\tiny{F}}}$, which is located inside the insulating gap of the alkyl chain. For this reason, all the $k_\alpha$'s have finite imaginary parts, which are always taken with positive sign. For each such $k_\alpha$, there is an evanescent Bloch solution $\psi_{k_\alpha}(\brr)$ exponentially decaying to zero when $z$ goes to positive infinity, and an evanescent Bloch solution $\psi_{-k_\alpha}(\brr)$ exponentially decaying to zero when $z$ goes to negative infinity. The exponential decay factor of the evanescent solutions is given by Im[$k_\alpha$]. The alkyl chains conduct through all the evanescent channels, but for long chains the charge transport mainly occurs through the channel with minimum Im[$k_\alpha$], the contributions from the other channels being exponentially small. Therefore, we can focus on the channel with the lowest Im[$k_\alpha$] and for this reason we drop the index $\alpha$ from now on.  

\begin{figure}
  \includegraphics[width=7cm]{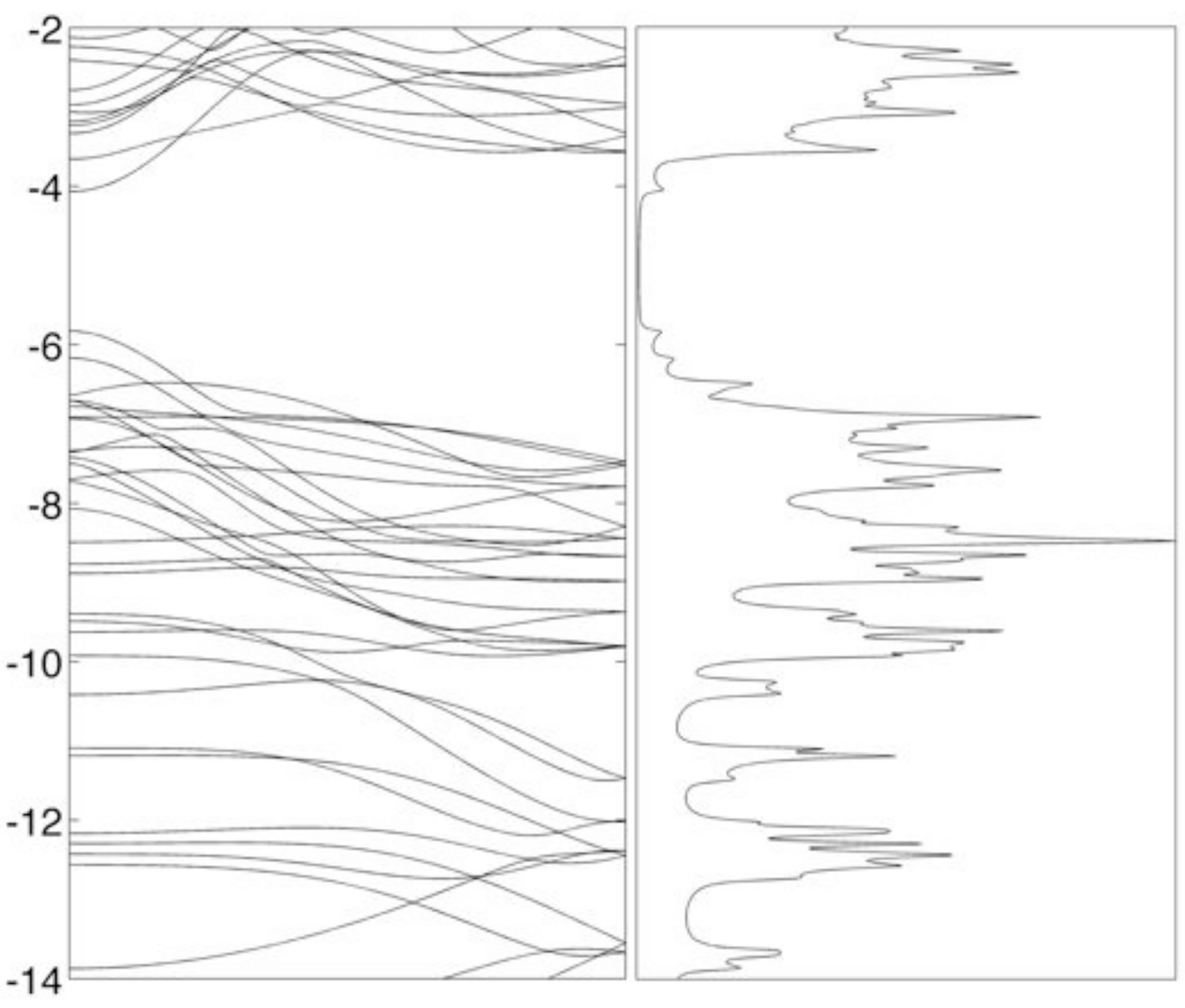}
  \caption{The band structure (left) and density of states (right) of the infinite Si nano-wire. The energy units are eV.}
 \label{leadsbands}
\end{figure}

To compute the transmission of the whole device, written in Eq.~\ref{trans}, one can place the normal planes $\brr_\bot$ and $\brr'_\bot$ in the middle of the device, in which case one only needs to understand how the evanescent channels reflect at the contacts on to the potentials $V_{\mbox{\tiny{L/R}}}(\brr)$. This has been accomplished in Refs.~\onlinecite{Prodan:2007qv} and \onlinecite{Prodan:2009cs} by using a Green function formalism. The asymptotic expression of $T(\epsilon)$, in the limit of long alkyl chains, was found to be:
\begin{equation}\label{trans}
T(\epsilon) = \Theta_{\mbox{\tiny{L}}} \Theta_{\mbox{\tiny{R}}}  e^{-2Im[k]L}.
\end{equation}
The $\Theta$ coefficients are given by:
\begin{equation}\label{thetal}
\begin{array}{c}
\Theta_{\mbox{\tiny{L}}}=\frac{2\pi}{W(\psi_k,\psi_{-k})} \int d \brr \int d \brr'  \times \medskip \\
\psi_{-k}(\brr)\Delta V_{\mbox{\tiny{L}}}(\brr)\rho_{\epsilon_F}(\brr,\brr')\Delta V_{\mbox{\tiny{L}}}(\brr')\psi_{-k}(\brr'),
\end{array}
\end{equation}
with $\brr$ and $\brr'$ measured from the left end of the chain. Similarly
\begin{equation}\label{thetar}
\begin{array}{c}
\Theta_{\mbox{\tiny{R}}}=\frac{2\pi}{W(\psi_k,\psi_{-k})} \int d \brr \int d \brr'  \times \medskip \\
\psi_{k}(\brr)\Delta V_{\mbox{\tiny{R}}}(\brr)\rho_{\epsilon_F}(\brr,\brr')\Delta V_{\mbox{\tiny{R}}}(\brr')\psi_{k}(\brr'),
\end{array}
\end{equation}
with $\brr$ and $\brr'$ measured from the right end of the chain.
In both expressions, $\rho_{\epsilon_F}(\brr,\brr')$ is the spectral operator:
\begin{equation}\label{spectral}
\rho_{\epsilon_F}(\brr,\brr')=\sum\limits_\epsilon \frac{ \delta/\pi}{(\epsilon-\epsilon_F)^2+\delta^2} \phi_\epsilon^*({\bf r}) \phi_\epsilon({\bf r}').
\end{equation}
The sum goes over all Kohn-Sham orbitals $\phi_\epsilon(\brr)$. The symbol $W(\phi,\psi)$ stands for the generalize Wronskian: 
\begin{equation}
W(\phi,\psi)=\int d \brr_\bot \ \phi(\brr)\overleftrightarrow{\partial_z} \psi(\brr).
\end{equation}
The expressions for $\Theta$ coefficients are independent of how one normalizes the evanescent Bloch functions.

\begin{figure}
\center
  \includegraphics[width=7cm]{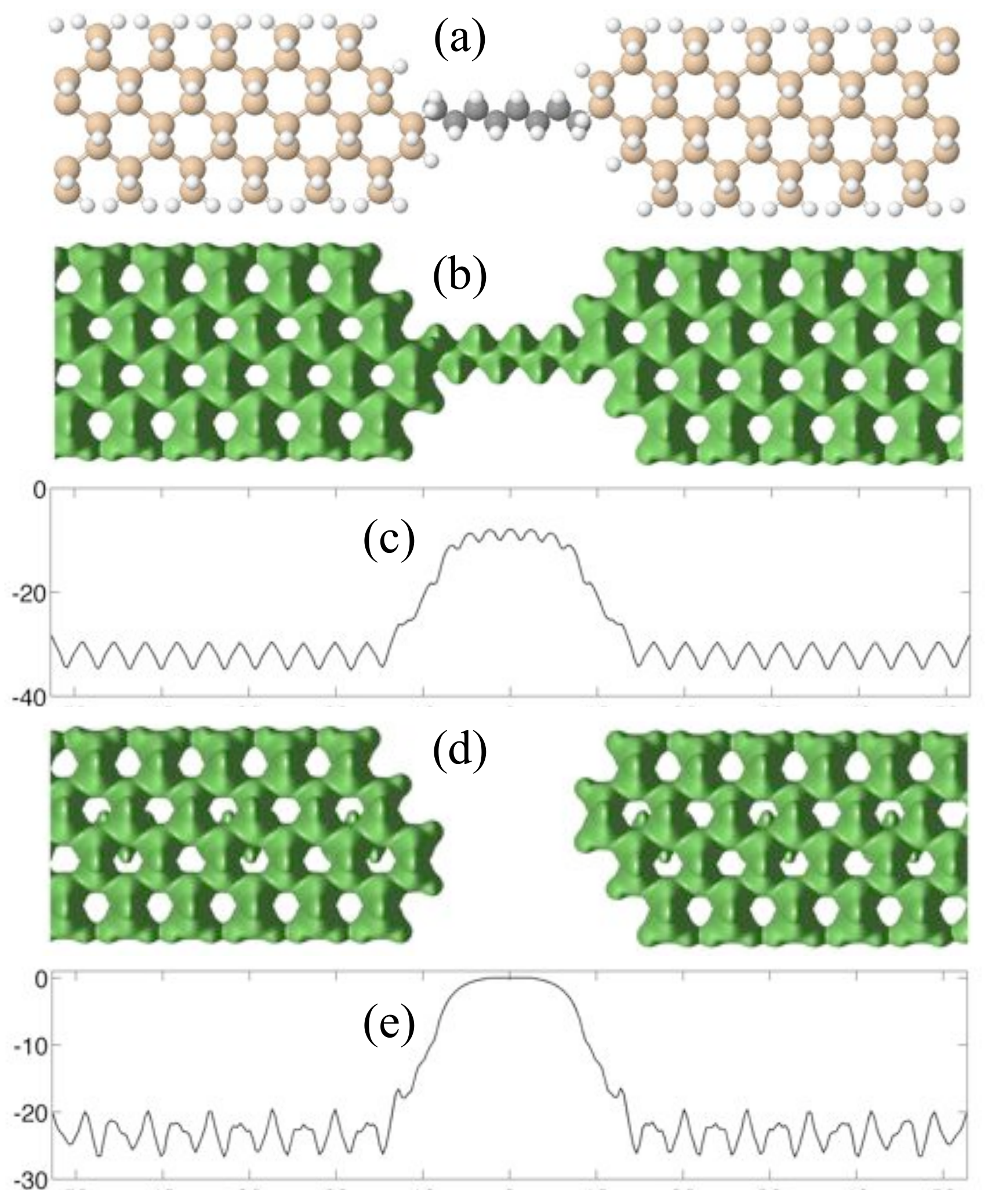}\\
  \caption{(Color online) (a) The atomic configuration of the device. b) An iso-surface of $V_{\mbox{\tiny{eff}}}$. c) The xy average of $V_{\mbox{\tiny{eff}}}$. d) An iso-surface of $\Delta V$. e) The xy average of $\Delta V$. The energy units in all panels are in eV.}
 \label{DiffPot}
\end{figure}

A closer look at the integrands of Eqs.~\ref{thetal} and \ref{thetar} reveals that $\psi_{-k}(\brr)\Delta V_{\mbox{\tiny{L}}}(\brr)$ and $\psi_{k}(\brr)\Delta V_{\mbox{\tiny{R}}}(\brr)$ are exponentially localized near the contacts. Therefore, to evaluate the $\Theta$ coefficients, one only needs a converged spectral operator near the contacts, which can be obtained from a standard electronic structure calculation which includes long enough leads.

The asymptotic expression of $G$ becomes virtually exact for alkyls containing more than 4 methyl groups.\cite{Prodan:2008by} The simplification brought in by this expression allows us to compute the tunneling conductances of the extremely large molecular devices presented in this paper without truncating the Hilbert space. We use the same number of basis functions (better said grid points) in the transport calculation as in the self-consistent calculation, which for one device is larger than 2.1$\times$10$^6$. Most of the transport calculations completed so far use a number of basis functions that is orders of magnitude smaller than the ones used in this paper.

Besides the computational advantages, the analytic expression for $G$ allows us to get an unprecedented inside look into the transport characteristics of the devices. Since the expressions for the $\Theta$ coefficients are basically overlap integrals, each quantity inside the integrands reveals one specific aspect. The evanescent Bloch functions allows us to image the path that is important for tunneling transport. A plot of $\psi_{-k}(\brr)\Delta V_{\mbox{\tiny{L}}}(\brr)$ and $\psi_{k}(\brr)\Delta V_{\mbox{\tiny{R}}}(\brr)$ reveals the localization of the contact conductance, allowing us to determine which layers inside the leads are important for the tunneling transport.

The formalism is applied to devices made of alkyl chains connected to Si nano-wires, as illustrated in Fig.~\ref{setup}. Various alkyl chains and Si nano-wires will be considered. All the broken Si bonds are passivated with hydrogen and all the bonds are kept in a perfect tetrahedral configuration. The Si atoms are arranged in the bulk configuration, the S-H bond length is fixed at 1.49\ \AA \ and the Si-C bond at 1.75\ \AA. For the alkyl chains, the C-C bond length is fixed at 1.54 \ \AA \ and the C-H bond length at 1.1\ \AA.

The calculations are performed in a periodic supercell containing the alkyl chain and the Si nano-wires as illustrated in Fig.~\ref{setup}. The equilibrium self-consistent Kohn-Sham calculations, the real and complex band structure calculations and the non-equilibrium transport calculations were performed with a real space, pseudopotential code based on finite differences. The same code was used for the calculations reported in Ref.~\onlinecite{Prodan:2008by} and \onlinecite{Prodan:2009cs}. We adopted a 5-point (in each direction) finite difference approximation for the kinetic energy operator, and used a uniform space grid of spacing 0.19 \AA, sufficient for a good convergence of the electronic band structure. This grid is commensurate with the unit cell of the periodic alkyl chain, which is the reference system in our transport calculations. The largest device discussed in this paper requires the staggering number 81$\times$81$\times$329 of grid points and contains 1092 electrons. The lateral grid size is kept constant at 81$\times$81 grid points or 15.4\AA$\times$15.4\AA. 

\begin{figure*}
  \includegraphics[width=18cm]{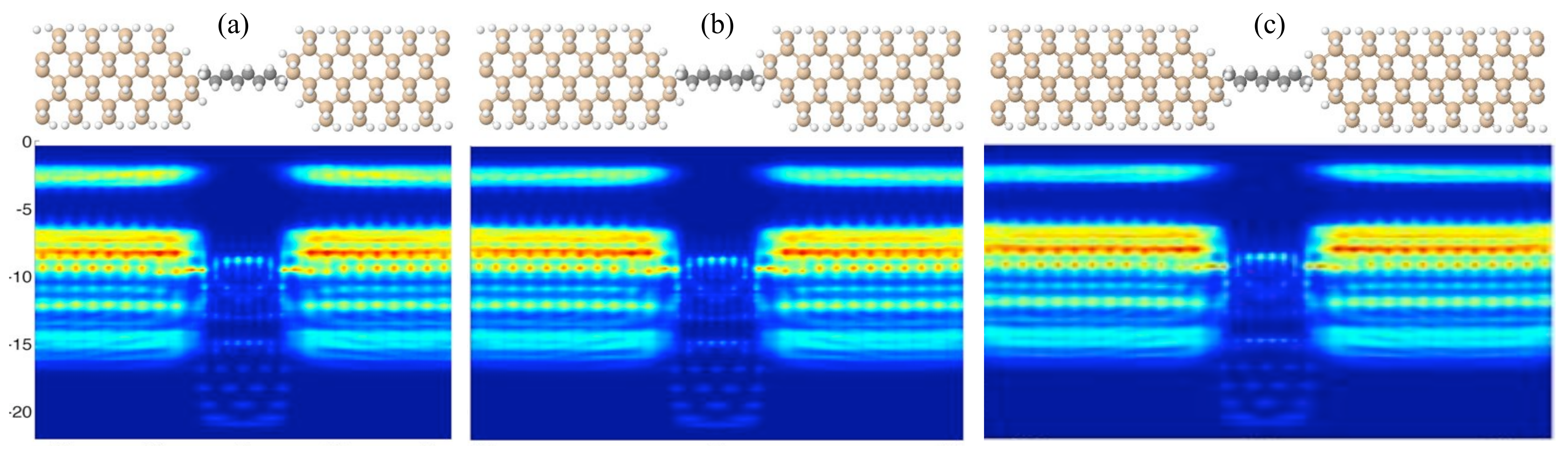}
  \caption{(Color online) The atomic configurations of the three molecular devices used to investigate the thermodynamic limit and the corresponding xy average of the local density of states. The energy unit is 1 eV.}
 \label{devices1}
\end{figure*}

 We adopted the Local Density Approximation (LDA) for exchange and correlation using the Perdew-Zunger (PZ) \cite{Perdew:1981gb} interpolation of the numerical electron-gas data of Ceperley and Alder. \cite{Ceperley:1980eu} We used Troullier-Martin norm-conserving pseudo-potentials \cite{Troullier:1991ys} for all the atomic species. The pseudopotentials for C and Si atoms had distinct {\it s} and {\it p} components and we took the {\it p} pseudo-potential as the local reference. Purely local pseudopotentials were used for the H atoms.

\section{The semi-conducting electrodes}

Although the present work was motivated by the transport measurements on self-assembled alkyls on Si(111) surface, we cannot approach these systems at the moment. Due to the tilting of the alkyl chains relative to the Si(111) surface, a computation on the alkyl mononlayer will require a non-rectangular suppercell. To avoid such computational complication, we chose to represent the Si electrodes by Si nano-wires grown along the [110] direction. The Si nano-wires are cut along the [111] surface where the alkyl chain is then attached. In principle, by taking the diameter of the wire larger and larger, we can model a low density alkyl monolayer. But on the other hand, since Si nano-wires have been fabricated for some time, the devices discussed in this work could very well be fabricated in the near future.  For example, free standing silicon wires grown along directions like [100], [110] or [112] have been already observed experimentally.\cite{Ma:2003gb,Holmes:2000mb}  Their electronic properties and their potential for novel electronic devices were discussed in Refs.~\onlinecite{Zhao:2003ud,Zhao:2004zp,Scheel:2005vl,Markussen:2006sf,Rurali:2007jt}. 

In this work we consider the thinnest [110] hydrogen passivated Si nano-wires. The unit cell of the wire contains 16 Si and 12 H atoms. The atomic configuration of the Si nano-wire is illustrated in Fig.~\ref{siwire} from various view angles. The Si atoms are fixed at the bulk positions as previously discussed. This model structure is a fairly accurate representation of the relaxed configurations, the difference between the two being less than 1\%.\cite{Scheel:2005vl} 

As already pointed out in Ref.~\onlinecite{Zhao:2004zp}, the confinement of the electrons in thin Si wires leads to an electronic structure that is quite different from the bulk Si. In Fig.~\ref{leadsbands} we show the band structure of our Si nano-wire and the corresponding density of states. The band structure and the density of states near the band edges in Fig.~\ref{leadsbands} is similar those reported by previous calculation.\cite{Scheel:2005vl} The nano-wire displays a direct gap of 1.75 eV, in line with the previous theoretical works.\cite{Scheel:2005vl} As one can see in Fig.~\ref{leadsbands}, the insulating gap is determined by highly dispersive bands. This bands contribute very little to the total density of states but, as we shall see, they have large effect on the transport. This particularity of the Si nano-wires makes the transport problem more interesting, but at the same time more difficult because it slows the convergence with the leads size.

\begin{figure}
\center
  \includegraphics[width=8cm]{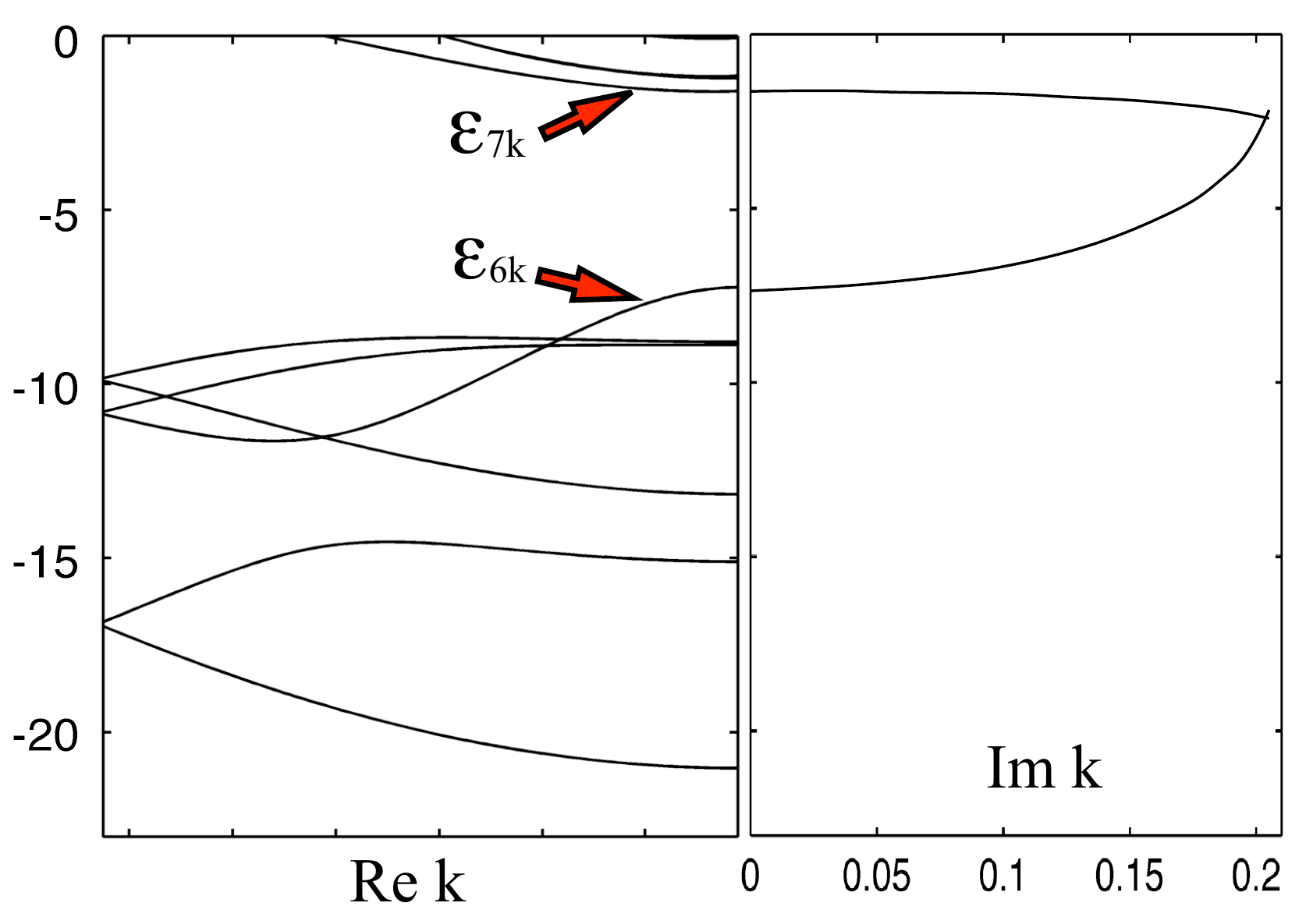}\\
  \caption{Real (right panel) and complex (left panel) band structures of the periodic potential $V_0$ corresponding to the device shown in Fig.~\ref{DiffPot}. Only the complex band with smallest Im[$k$] is shown. The energy unit is 1 eV and the unit for Im[$k$] is 1/Bohr.}
 \label{bands}
\end{figure}

\section{Electronic Structure: General considerations}  

First, let us discuss the periodicity of the effective potential inside the alkyl chain, because this is the main assumption of our theory. The full effective potential has local and non-local parts. Since we work with norm-conserving pseudopotentials, the non-local part of the potential is not changed during the self-consistent calculation. Therefore, the non-local part of the effective potential remains perfectly periodic inside the alkyl chain. In Fig.~\ref{DiffPot}(b-c) we plot the local part of the effective potential for the entire molecular device, shown in Fig.~\ref{DiffPot}(a). According to these graphs, the effective potential is periodic inside the alkyl chain, starting from the very first methyl group. The periodic potential $V_0$ is obtained by repeating the local and non-local parts of $V_{\mbox{\tiny{eff}}}$ inside the mid unit cell. If we subtract this $V_0$ from $V_{\mbox{\tiny{eff}}}$ we obtain the local potential difference $\Delta V$, which is plotted in Fig.~\ref{DiffPot}(d-e). $\Delta V$ rapidly decays to zero away from the contacts and stays at zero for most of the alkyl chain's length. By construction, the non-local part of $\Delta V$ is automatically zero on the alkyl chain. We recall that $\Delta V$ appears explicitely in the expression of the $\Theta$ coefficients, therefore Fig.~\ref{DiffPot} is also important for understanding the results on the conductance. The above conclusions hold true for all the electronic devices considered in this work.

The second issue we want to discuss is the fact that our molecular devices display a global insulating gap. This is evident in all the plots showing the local density of states (see for example Fig.~\ref{devices1}). The gap of the Si nano-wire is clean of electronic states because the broken Si bonds were carefully passivated with hydrogens. It is also important to notice that the gap is clean at the contacts, and this is again because we passivated the broken Si bonds resulting from clipping the Si nano-wires. Referring to Fig.~\ref{devices1}, one can distinctly notice the upper edge of the valence states for both Si and alkyl wires and the lower edge of the conduction states of the Si nano-wires, but only a faint mark of the conduction states edge of the alkyl chain can be observed. This is because we have computed only 75 un-occupied Kohn-Sham orbitals. Since we do not have a good description of the conduction bands, in our transport calculations we assume p-doped wires and therefore a Fermi level that is close to the valence bands.

The self-consistent calculations were performed at a finite temperature. In this case, we have a well defined Fermi level, but its position is irrelevant because it is highly sensitive to the presence of any impurity or defect, which will inherently occur in real devices. Because we include a fairly large vacuum region around the devices, the vacuum level is well defined and we can reference the electronic states from it. This is done throughout the paper.

The real and complex band structures of the periodic potential $V_0$ for the largest device of Fig.~\ref{devices1} are shown in Fig.~\ref{bands}. The complex bands remain almost unchanged when comparing one device to another. Apart from a rigid shift, both structures in Fig.~\ref{bands} are similar to those reported in Ref.~\onlinecite{Picaud:2003qf} for free standing alkyl chains. This tells us that, apart from a rigid shift coming from band alignment, $V_0$ is similar to the effective potential of free standing chains. In Fig.~\ref{bands} we show only the complex bands that are relevant to the tunneling. Note that the upper edge of the insulating gap practically terminates into the vacuum. This is a well known shortcoming of LDA, which places the bands too high in energy, though the shape of the bands is quite well reproduced, when compared to the experimentally resolved bands.\cite{Miao:1996fv}  For this reason, LDA gives a much smaller gap than the experimentally measured one. Nevertheless, if one follows the complex band originating from the molecular valence band,\cite{Picaud:2003qf} one will find that it connects to a band that is approximately 9 eV higher in energy, an energy difference that is close to the experimentally measured insulating gap of the alkyl chains. For this reason, the complex band shown in Fig.~\ref{bands} is expected to have the right shape, thus leading to correct values of $\beta$, provided the Fermi level is correctly predicted.

 \begin{figure}
  \includegraphics[width=8.6cm]{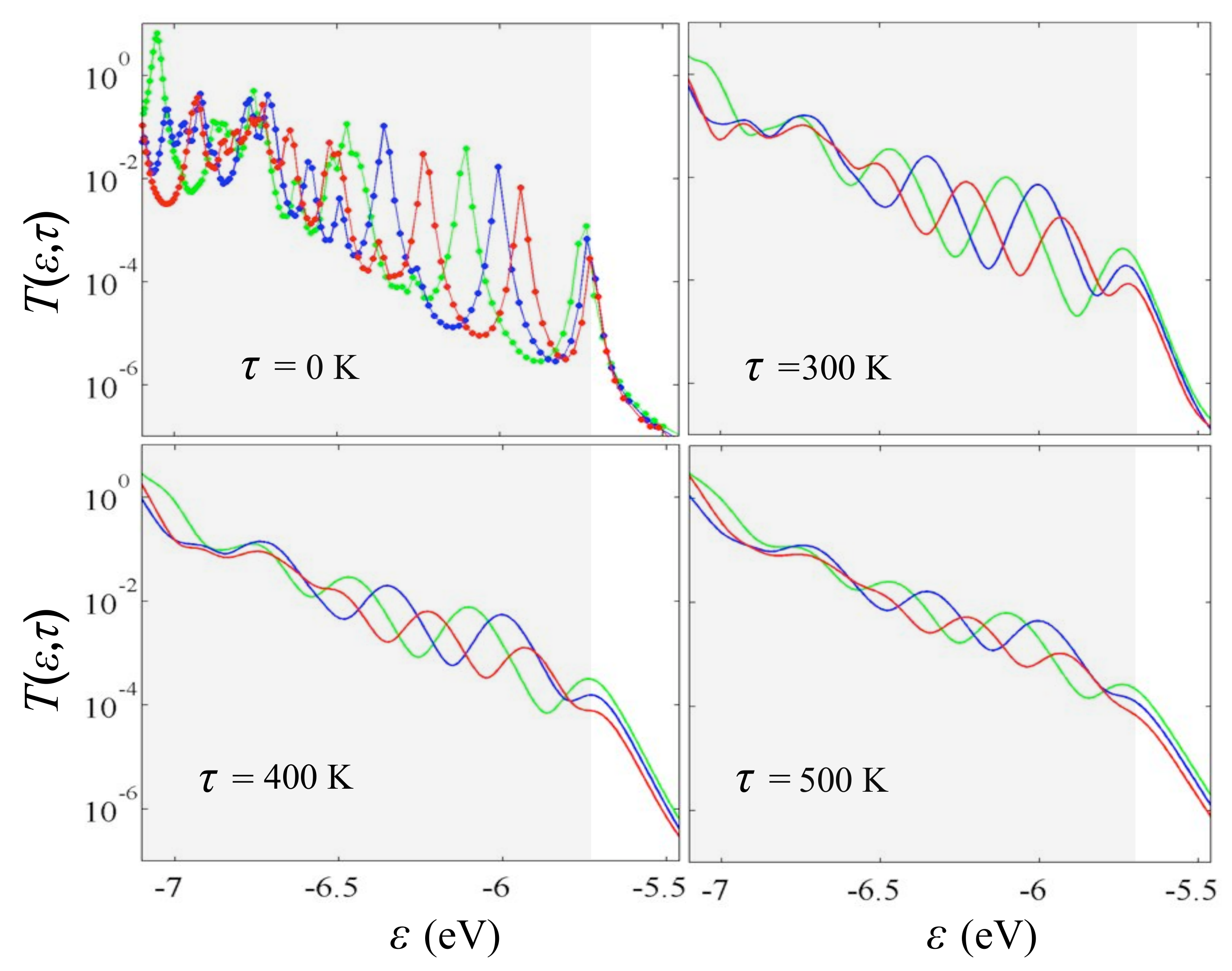}\\
  \caption{(Color online) Plots of the finite temperature transmission computed at 0, 300, 400 and 500 K, for the three device shown in Fig.~\ref{devices1}. The colors green, blue and red are used for devices (a), (b) and (c), respectively. The shaded region indicates the energy range of the valence states of the devices. The Fermi level of the device is located to the right of this region.}
  \label{TDLimit}
\end{figure}

\section{Thermodynamic limit}

In this section we discuss the convergence of our results with the size of the leads, which is always an important issue when working with finite systems.\cite{Verstraete:2009sw} For our case, we studied three devices with finite Si nano-wire leads containing a total of 9, 11 and 13 unit cells. These molecular devices are shown in Fig.~\ref{devices1}. In the same figure, below the atomic structures, we show the corresponding xy average local density of states: $\rho_{\mbox{\tiny{av}}}(z,\epsilon)$=$\int \rho_\epsilon(x,y,z)dxdy$. The plots give a color map of $\rho_{\mbox{av}}(z,\epsilon)$ in the plane of energy $\epsilon$ (in eV) and of position along the device. For the density of states, we used a smearing factor $\delta$ of $0.2$ eV. At the scale of these plots, the local density of states look completely converged, the plots showing no difference when passing from one device to the other. It is also important to point out that the edges of the valence and conduction states for both Si nano-wire and alkyl chain do not change from one device to the next.

Smearing factors of 0.1 to 0.2 eV are typical in transport calculations of $T(\epsilon)$.\cite{Tomfohr:2004ve} Here, however, since we perform finite temperature calculations, we do not want the $\delta$ smearing to interfere with the physical temperature smearing. Therefore, we must use a much smaller $\delta$ than in the typical transport calculations.  For example, the transport calculations at 300 K would require a $\delta$ less than $0.026$ eV and at 500 K a $\delta$ less than $0.043$ eV. On average, the energy level spacing in our devices is much less than 0.02 eV. Unfortunately, the transport is very sensitive to the electronic structure near the top of the valence band (assuming p-doped wires). This structure is determined by a very dispersive band therefore the spacing between the energy levels near the top of the valence band becomes much larger. For example, the average level spacing for the first five levels from the top of the valence band is 0.11 eV. 

  \begin{figure}
  \includegraphics[width=8cm]{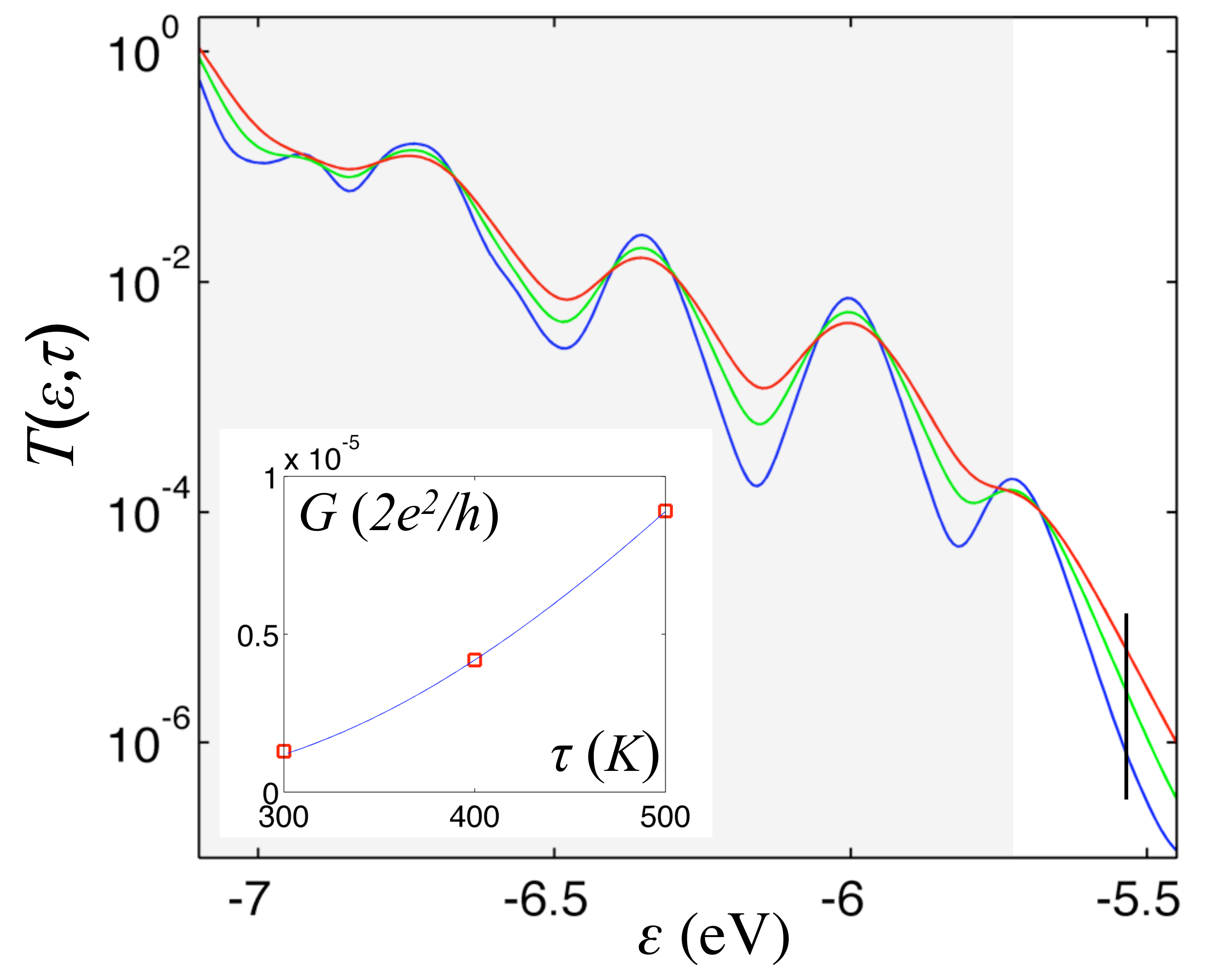}\\
  \caption{(Color online) Plots of the finite temperature transmission computed at 300 K (blue), 400 K (gree) and 500 K (red). The inset shows the values of the linear conductance at these three temperatures, with the Fermi level assumed to be position at the mark shown in the main graph. The continuous line in the inset is a plot of the function $g e^{\frac{\epsilon_{\text{\tiny{F}}}-\epsilon_{\text{\tiny{V}}}}{k\tau}}$, with $g$ adjusted till the best fit was obtained. The shaded region indicates the energy range of the valence states of the devices. The Fermi level of the device is located to the right of this region.}
 \label{TDep}
\end{figure}

\begin{figure*}
  \includegraphics[width=18cm]{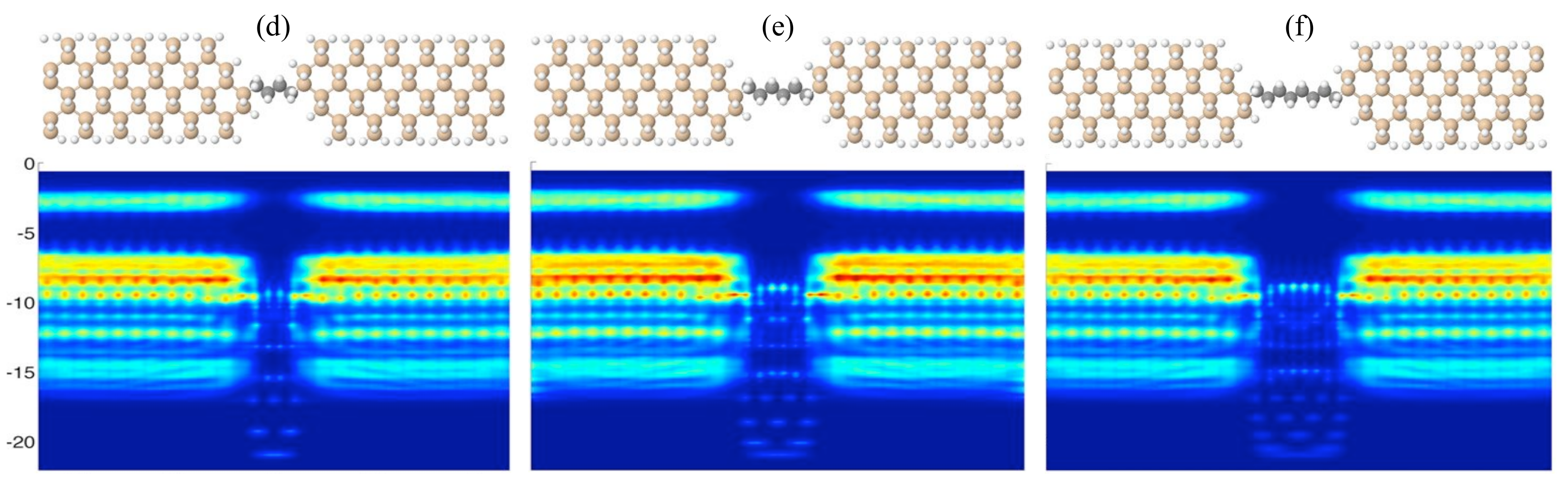}
  \caption{(Color online) The atomic configurations of the three molecular devices used to investigate the dependence on the number of monomers of the alkyl chain and the corresponding xy average of the local density of states. The energy unit is 1 eV.}
 \label{devices2}
\end{figure*}

The above criteria are only qualitative and we should look directly at the conductance results, which are shown in Fig.~\ref{TDLimit} for various temperatures. The 0 K plot shows the transmission $T(\epsilon)$ computed with a $\delta$ smearing of 0.01 eV, which was kept at this value throughout the paper. The next plots show the convolution of $T(\epsilon)$ with the first derivative of the Fermi-Dirac distribution at different temperatures $\tau$:
\begin{equation}\label{ttrans}
T(\epsilon,\tau)=\int  [-f'_{\text{\tiny{FD}}}(\epsilon-\epsilon';\tau)] \ T(\epsilon')  d\epsilon'.
\end{equation}
We call this quantity the finite temperature transmission. The finite temperature conductance of the devices is then given by $T(\epsilon,\tau)|_{\epsilon=\epsilon_F}$. $T(\epsilon,\tau)$ can be accurately computed only in a limited energy window. The lower limit of this energy window is determined by the edge of the $V_0$ potential's valence band, because our theory cannot be applied below this energy limit. The upper limit of the energy window is determined by the energy where the spurious Lorentian tails seen in the first panel of Fig.~\ref{TDLimit} become larger than the physical tails of the Fermi-Dirac distributions. Beyond this point, $T(\epsilon,\tau)$ decays as a Lorentian, a spurious effect due to the finite $\delta$ smearing used in the calculations.

Looking at $T(\epsilon,\tau)$ for different temperatures, we see the thermodynamic limit being accelerated as the temperature increases. The size dependence becomes quite small in the entire energy window, if one looks at the plots for 500 K.  But the most important fact is that $T(\epsilon,\tau)$ shows almost no size dependence in the energy window above the valence states, for all three temperatures.

\section{Dependence on the Temperature}

In this section we explore the behavior of the tunneling conductance for different temperatures. In Fig.~\ref{TDep} we show $T(\epsilon,\tau)$ at $\tau$=300, 400 and 500 K for the largest device shown in Fig.~\ref{devices1}.  For energies below the insulating gap, the temperature effects would be negligible if we had a converged density of states. In our case, we do see a temperature effect below the insulating gap and, as expected, $T(\epsilon,\tau)$ becomes smoother as the temperature increases. Inside the insulating gap, we see large variations of $T(\epsilon,\tau)$ with the temperature and these variations are physical. It is important to observe that the  temperature dependence becomes weaker until it disappears as the Fermi level gets closer and closer to $\epsilon_{\text{\tiny{V}}}$, the edge of the valence band. We will come back to this issue, after we map the dependence of conductance on the number of monomers, when we can put together the scenario describing the experimental observations.

To see how the conductance itself behaves with the temperature, we kept the Fermi level pinned at the same value as we varied the temperature. This value was chosen to be -5.55 eV, about 0.16 eV above the leads' valence band. From the main plot of Fig.~\ref{TDep}, we collected the three values of the conductance corresponding to this $\epsilon_{\text{\tiny{F}}}$, which are shown in the inset of the same figure. The continuous line in the inset of Fig.~\ref{TDep} represents the function $g e^{\frac{\epsilon_{\text{\tiny{F}}}-\epsilon_{\text{\tiny{V}}}}{k\tau}}$, which describes quite well the computed values for the conductance. This is precisely what the transport measurements for self-assembled alkyls on Si surface show.

Few final notes for the Section are in place. The present calculations use the  finite temperature DFT formalism to account for temperature effects on the electronic structure. However, since here we are discussing room temperatures ($\pm$ 200 K), the electronic structure is practically unchanged from that at 0 K. The issue of electron-phonon scattering inside the device could be raised. If the effect would have had any observable effects, the behavior of $G$ with the length of the insulating chain would deviate from the purely exponential tunneling decay behavior. But we know from the extremely well controlled experiments on amine-linked alkyl chains on gold leads,\cite{Venkataraman:2006lq} that the exponential tunneling  decay behavior is obeyed quite strictly [see also the discussion in Refs.~\onlinecite{Nitzan:2003rr} and \onlinecite{Galperin:2007cr}]. There is no reason for things to be different when the alkyl chains are attached to Si wires and therefore we conclude that the electron-phonon scattering inside the device is negligible. It will definitely become an issue if one is concerned with the resistance of the long Si leads, which we are not. 

Another concerned could be structural molecular changes induced by temperature variations. The transport experiments,\cite{Salomon:2003gd,A.-Salomon:2005kx,Nesher:2006ys,Seitz:2006rw,Seitz:2007dk,Salomon:2007qr,Thieblemont:2008hl,Aswal:2006rm} were never able to resolve any effects coming from possible structural changes because, even if they occur, they are dwarfed by the large changes in the conductance already captured and discussed in Fig.~\ref{TDep}. 

\section{Dependence on the number of monomers}

For this part of our study, we work with the devices shown in Fig.~\ref{devices2}, containing 4, 6 and 8 methyl groups. Similarly to Fig.~\ref{devices1}, this figure illustrates the xy averaged local density of states for the three devices. The plots show a good alignment of the energy levels when comparing one device to the next. Also, the plots show almost no changes in local density of states near the contacts and inside the Si nano-wires, a fact that strongly suggests that the asymptotic regime was reached. 

\begin{figure}
  \includegraphics[width=8.6cm]{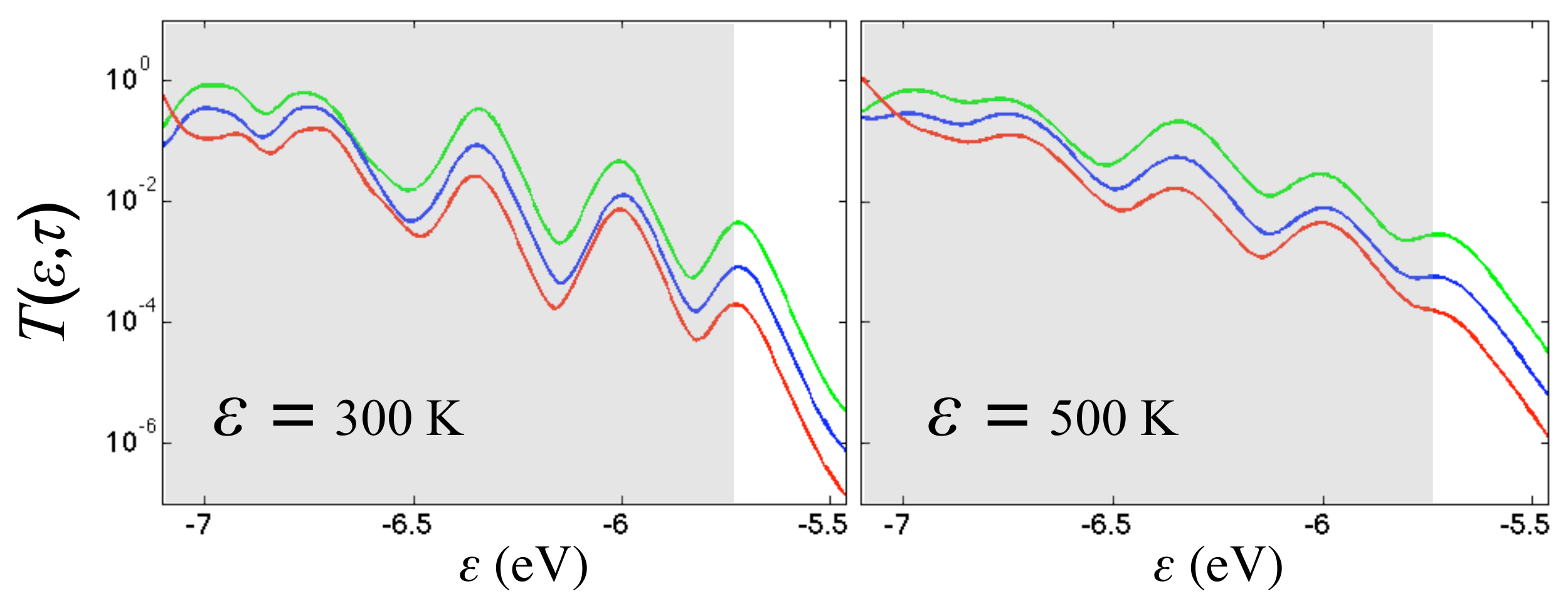}\\
  \caption{(Color online) Plots of the finite temperature transmission computed at 300 K and 500 K for the three devices shown in Fig.~\ref{devices2}. The colors green, blue and red are used for devices (d), (e) and (f), respectively. The shaded region indicates the energy range of the valence states of the devices. The Fermi level of the device is located to the right of this region.}
 \label{LDep}
\end{figure}

We computed the finite temperature transmission for these devices at different temperatures and the results are shown in Fig.~\ref{LDep}. According to this data, if we fix our attention to one arbitrary energy above the valence band, the transmission $T(\epsilon,\tau)$ decreases exponentially with the number $N$ of monomers as $T(\epsilon,\tau)\propto e^{-\beta N}$. Interestingly, $\beta$ is practically independent of temperature. Since the expression for $T(\epsilon,\tau)$ in Eq.~\ref{ttrans} is just the convolution of the zero temperature $T(\epsilon')$ and the derivative of the Fermi Dirac distribution, $T(\epsilon,\tau)$ is primarily determined by the $\epsilon'$ that are close to the upper edge of the valence band. The contributions from the $\epsilon'$ that are deeper into the valence band are exponentially suppressed. Therefore, we can not only understand the above observation, but we can actually tell what determines $\beta$, something that eluded scientists for some time. For energies above the valence band, $T(\epsilon,\tau)$ decays exponentially with the number of monomers and the exponential decay constant $\beta$ is temperature independent and given by 2Im[$k$]$b$ (see Eq.~\ref{trans}) with Im[$k$] evaluated at the top of the valence band. This is an important observation because it tells us that the tunneling decay constant in devices with semi-conducting leads is not determined by the Fermi level but rather by the edge of the valence band, for p-doped leads, and by the edge of the conduction band for n-doped leads. The present calculations give $\beta$=0.73.

\section{Bridging with the experiment}

\begin{figure}
  \includegraphics[width=6cm]{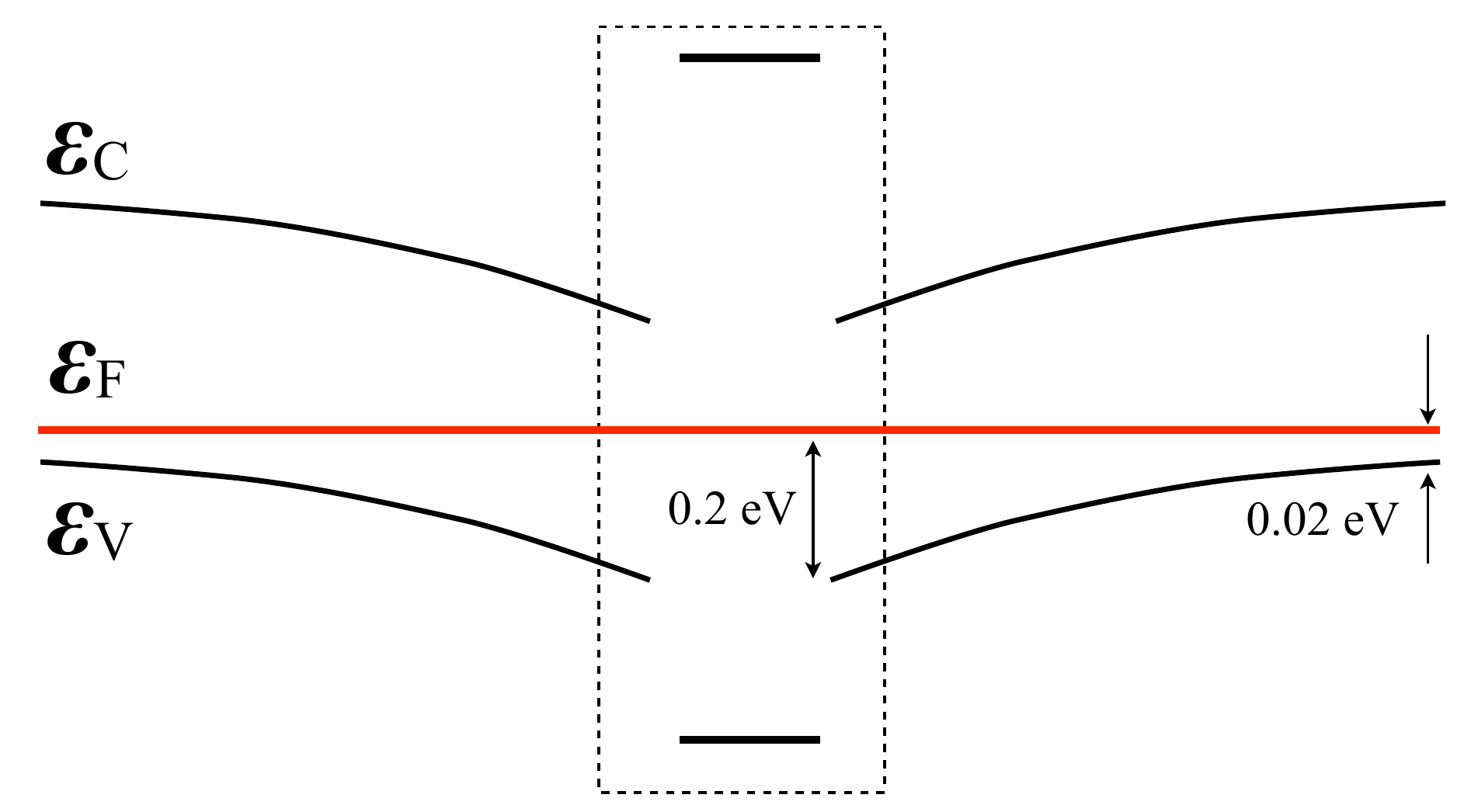}\\
  \caption{(Color online) Qualitative illustration of the local density of states for a real molecular device connected to a p-doped Si electrode.  Typical energy values are also shown. The rectangle drawn with a dotted line represents the portion of the device that can we capture in our atomistic simulations.}
 \label{rdevice}
\end{figure}

So far we have been careful not to place a firm Fermi level in our figures and we showed only results for the transmission rather conductance (with one exception), because determining the position of the Fermi level is beyond our atomistic simulations. In Fig.~\ref{rdevice} we show a qualitative diagram of the local density of states in real devices, assuming p-doped Si electrodes. The box shown in this figure represents the part of the device that we can actually model with our atomistic simulations. In real devices, the position of the Fermi level is determined by the impurity doping,\cite{Nesher:2006ys,Seitz:2007dk} which is impossible to simulate from first principles. Moreover, the doping triggers a band bending\cite{Salomon:2007qr} that is also impossible to capture with any purely atomistic simulation. Therefore, $\epsilon_{\mbox{\tiny{F}}}$$-$$\epsilon_{\mbox{\tiny{V}}}$ cannot be predicted by our calculations. We also want to mention that our transport simulations exclude the tunneling through the Schottky barrier. However, this part is thoroughly understood and can be dealt with by well established theoretical treatments.\cite{Thieblemont:2008hl} 

In most measurements involving semi-conducting leads,\cite{Salomon:2003gd,A.-Salomon:2005kx,Nesher:2006ys,Seitz:2006rw,Seitz:2007dk,Salomon:2007qr,Thieblemont:2008hl} the experimental I-V curves display two distinct regions corresponding to thermionic and tunneling regimes. For example in Ref.~\onlinecite{A.-Salomon:2005kx}, the thermionic regime starts from zero bias potential and ends at about 0.6 eV bias potential.  It was argued that the main effect here comes from the flattening of the band bending as the bias is applied, bringing the alkyl chain's gap edge closer to the Fermi level (see Fig.~\ref{rdevice}). This part of the I-V curves was found to be sensitive to temperature and only slightly sensitive to the alkyl chain's length. Beyond the 0.6 V bias potential, the I-V curves display a linear behavior, with a slope that is highly dependent on the alkyl chain's length and almost independent of temperature. It was argued that bands are flat in this regime and that the I-V curves are determined by the tunneling through the alkyl chain. Almost as a rule, it was observed that the band bending in p-doped semiconductors is small and the tunneling regime starts from very low bias potentials.\cite{Salomon:2007qr}

\begin{figure}
  \includegraphics[width=8cm]{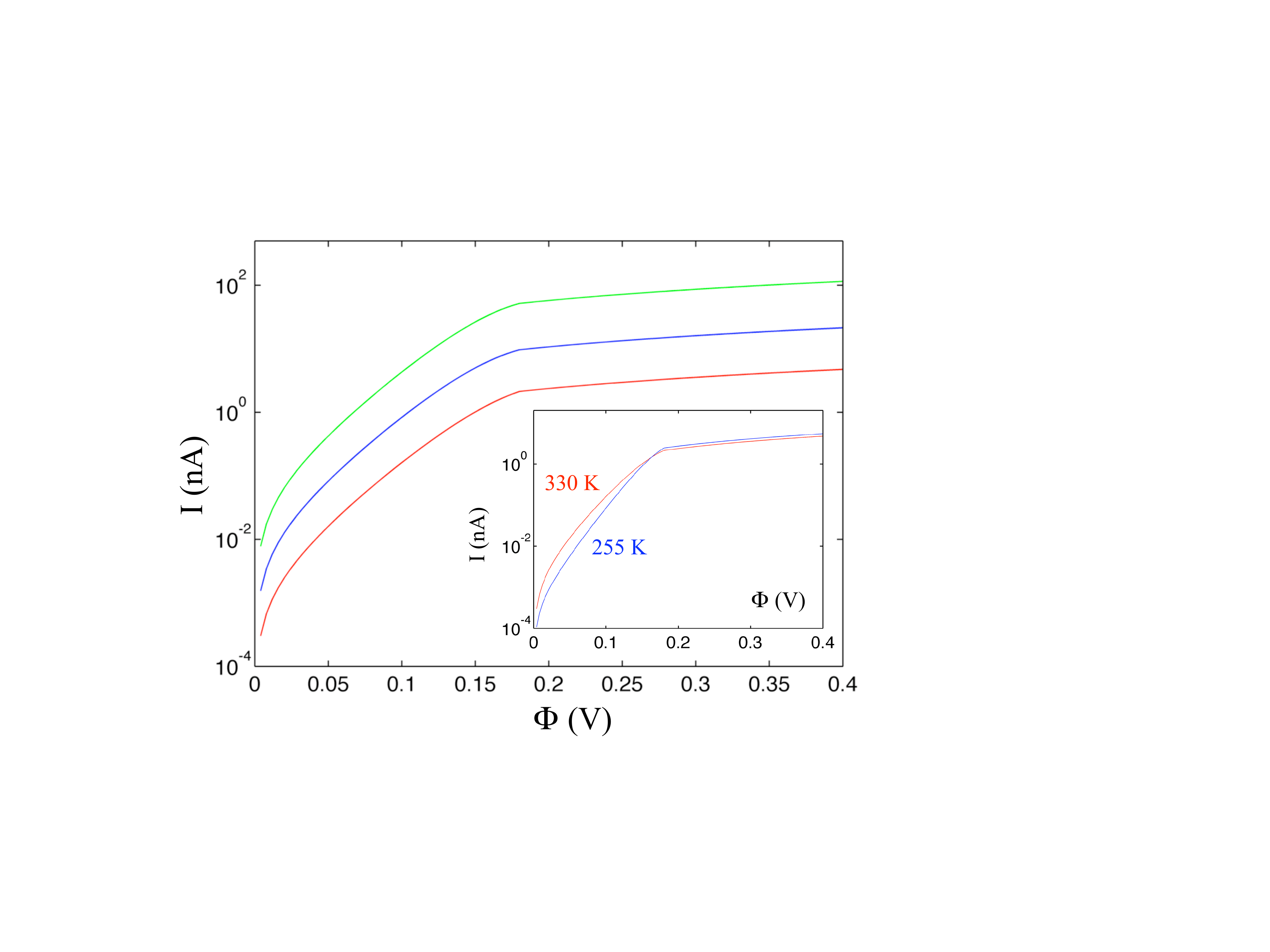}\\
  \caption{(Color online) The predicted I-V curves for the devices (d), (e) and (f) computed at 330 K. The inset shows the predicted I-V curve computed at 330K and 255 K as in Ref.~\onlinecite{A.-Salomon:2005kx}.}
 \label{IV}
\end{figure}

We argue that our finite temperature expression for the tunneling conductance can capture both regimes. We use our shorter alkyl chains to make qualitative contact with the experimental data mentioned above. To mimic the band flattening with the applied bias potential $\Phi$, we fix the valence and conduction bands but consider a bias potential dependent Fermi level (in eV): $\epsilon_{\mbox{\tiny{F}}}(\Phi)$=$\epsilon_{\mbox{\tiny{V}}}$+0.2$ -$$ \Phi$ for $\Phi$$<$0.18 eV and $\epsilon_{\mbox{\tiny{F}}}(\Phi)$=$\epsilon_{\mbox{\tiny{V}}}$+0.02 otherwise. This is consistent with the numbers shown in Fig.~\ref{rdevice} and assumes a band flattening that is linear with $\Phi$. In reality, the band flattening will take place only on one side of the device, but we do not have the means to account for that fact at the moment. Only relatively small bias potentials will be considered so that we can apply the linear response theory on our molecular devices. In Fig.~\ref{IV} we generate the I-V curves for the devices shown in Fig.~\ref{devices2}, by using the formula $I(\Phi)$=$G(\Phi)$$ \cdot$$ \Phi$, where the bias potential dependence in $G(\Phi)$ is induced by $\epsilon_{\mbox{\tiny{F}}}(\Phi)$. The expression for the current assumes that the potential $\Phi$ distributes entirely over the molecular device. For larger Schottky barriers we must take into the account that only a part of $\Phi$ is actually distributed over the device.

There is a strong qualitative agreement between our theoretical I-V curves and the experimental I-V curves for both n-doped and p-doped Si electrodes. The overall qualitative agreement is almost perfect with the experimental I-V curves for the p-doped electrodes of Ref.~\onlinecite{Salomon:2007qr}. For the n-doped Si electrodes,\cite{Thieblemont:2008hl} the Schottky barrier is much higher and evidently our I-V curves are missing the effects of the this barrier. Nevertheless, the entire shape of the experimental I-V curves\cite{Thieblemont:2008hl} is well reproduced by our theoretical I-V curves. The insight of Fig.~\ref{IV} shows two I-V curves computed at the two different temperatures considered in Ref.~\onlinecite{A.-Salomon:2005kx}. These theoretical I-V curves and the experimental I-V curves from Ref.~\onlinecite{A.-Salomon:2005kx}, are in very good qualitative agreement.

\section{Insight into the transport properties}

Let us briefly describe how we computed the zero temperature conductance transmission. We limit the discussion of this Section to the device (f) of Fig.~\ref{devices2}. Given the particular complex band structure of the alkyl chains, the tunneling conductance is determined by just one complex band, the one with the smallest Im[$k$]. This complex band is shown in Fig.~\ref{bands}. The complex band was obtained by varying continuously Im[$k$] from 0 to its maximum value, while keeping Re[$k$]$=0$. For each complex value of $k$, the spectrum of the $k$ dependent Hamiltonian:
\begin{equation}
H_k=-({\bm \nabla}-ik{\bm e}_z)^2+V_0+e^{-ik(z-z')}V_{\mbox{\tiny{non-loc}}}({\bm r},{\bm r}'),
\end{equation}
with periodic boundary conditions at $z=\pm b/2$, was calculated and its eigenvalues ordered according to their real parts: Re[$\epsilon_{1k}$]$<$Re[$\epsilon_{2k}$]$<$ \ldots . We focused, in particular, on the 6th  and 7th eigenvalues $\epsilon_{6k}$ and $\epsilon_{7k}$ (which take real values, see Fig.~\ref{bands}) and their corresponding evanescent Bloch functions $\psi_{6k}$ and $\psi_{7k}$. When Im[$k$]=0, $\epsilon_{6k}$ and $\epsilon_{7k}$ coincide, respectively, with the top of the valence band and with the bottom of the conduction band of the periodic potential $V_0$.  By increasing Im[$k$], the two eigenvalues move towards each other until they meet and actually cross each other. At various values of Im[$k$], we evaluated Eq.~\ref{trans} for both $\epsilon$=$\epsilon_{6k}$ and $\epsilon$=$\epsilon_{7k}$, using the corresponding evanescent Bloch functions $\psi_{6k}$ and $\psi_{7k}$ to compute the $\Theta$ coefficients via formulas \ref{thetal} and \ref{thetar}. The spectral kernel was computed directly from the Kohn-Sham orbitals of the full device as previously explained. The smearing coefficient $\delta$ was fixed at 0.013 eV.

  \begin{figure}
  \includegraphics[width=8.6cm]{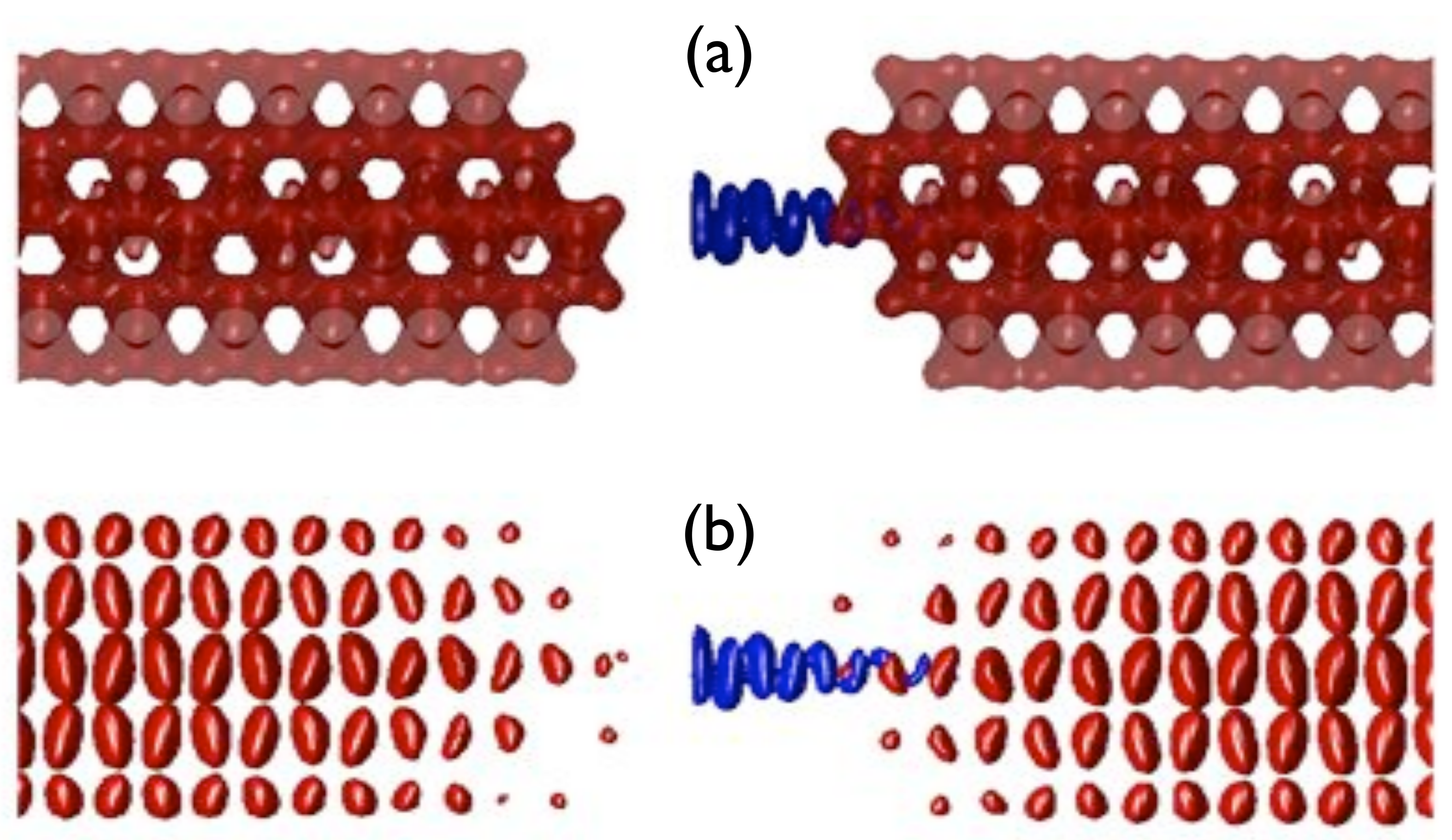}\\
  \caption{(Color online) Illustration of the overlap between the evanescent wavfunction $\psi_k({\bm r})$ and (a) $\Delta V$ and (b) $\rho_{\epsilon_{\mbox{\tiny{V}}}}(\brr)$.}
 \label{overlap}
\end{figure}

The analytic expressions of Eqs.~\ref{trans}, \ref{thetal} and \ref{thetar} allows us to point several key aspects of the tunneling transport of our devices. The $\Theta$ coefficients are given by the overlap of the evanescent wavefunctions of the the alkyl chain, the potential perturbation $\Delta V$ and the spectral function of the entire device. Therefore, we can gain valuable insight into the transport properties of our devices by mapping each of these three physical quantities and their overlap.

\begin{figure}
  \includegraphics[width=8.6cm]{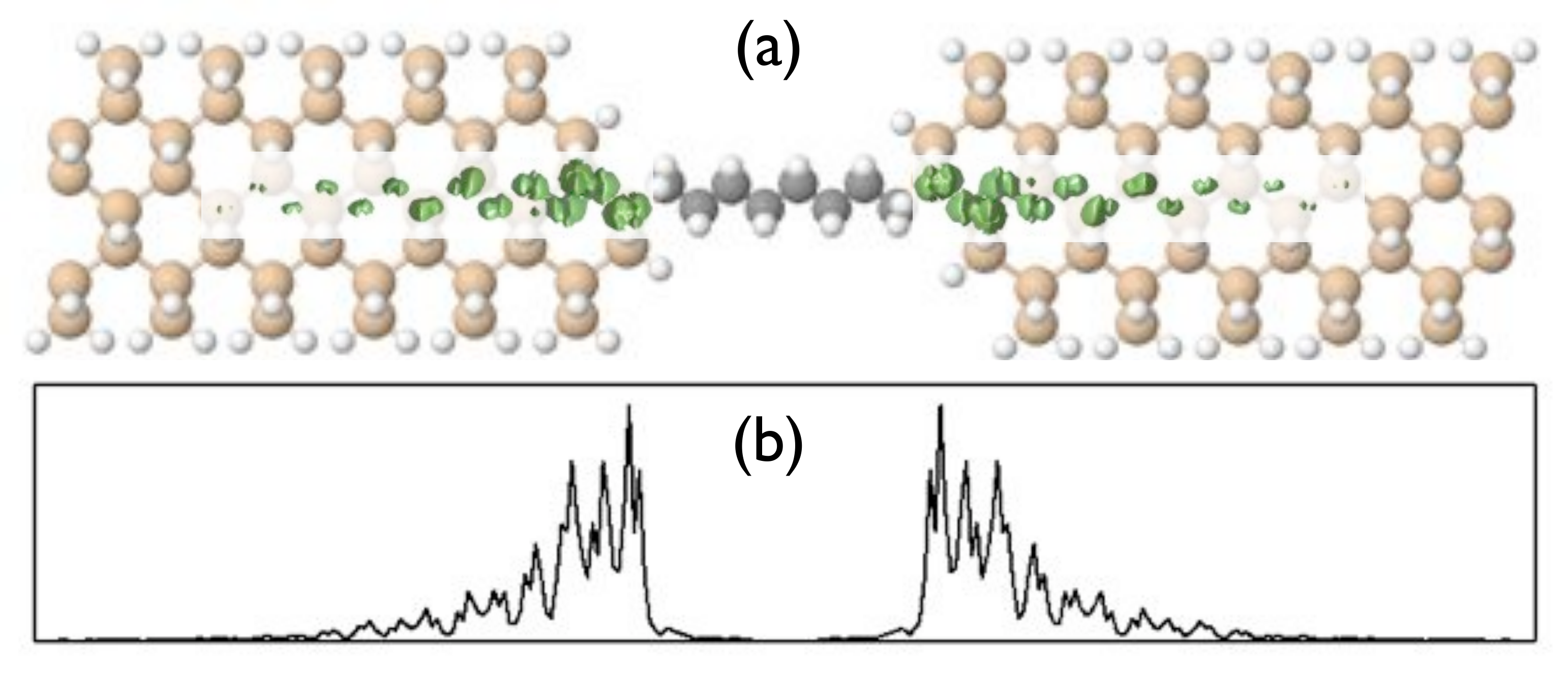}\\
  \caption{(Color online) (a) Iso-surface plots (in green) of $|\Psi_{\mbox{\tiny{L/R}}}({\bm r})|$ imposed over the atomic structure of the device. (b) The xy average of $|\Psi_{\mbox{\tiny{L/R}}}({\bm r})|$.}
 \label{psi}
\end{figure}

A plot of the averaged local density of states (i.e. the diagonal part of the spectral operator) was already given in Figs.~\ref{devices1} and \ref{devices2}. Fig.~\ref{DiffPot} illustrates the decomposition of $V_{\mbox{\tiny{eff}}}$ in terms of $V_0$+$\Delta V$. In Fig.~\ref{overlap}(a) we show the overlap between $\Delta V$ and the right evanescent function $\psi_k({\bm r})$ evaluated at $\epsilon_{\mbox{\tiny{V}}}$. This overlap is further analyzed in Fig.~\ref{psi}. Fig.~\ref{overlap}(b) shows the overlap of the evanescent wavefunction $\psi_k({\bm r})$ and the local density of states $\rho_{\epsilon_{\mbox{\tiny{V}}}}({\bm r})$ at $\epsilon_{\mbox{\tiny{V}}}$. The plot reveals a very strong overlap which explains why the top valence states give such a large contribution to the transmission, while they contribute so little to the density of states (see Fig.~\ref{leadsbands}). We should comment here that the density of states at $\epsilon_{\mbox{\tiny{V}}}$ is determined by the highly dispersive band seen at and near the top of the valence band in Fig.~\ref{leadsbands}. The character of the top valance bands in Si nano-wires have been thoroughly discussed in Ref.~\onlinecite{Zhao:2004zp}, and shown to originate from the lateral confinement of the electrons  (the atomic character remains the same as in the bulk Si). In fact, the top valence bands are quite well reproduced by a jellium model confined in a cylinder of same radius as the wires. As the lateral size of the Si nano-wire is increased, the band structure converges to that of bulk Si, therefore the top valance bands become less dispersive and the associated wavefunctions less localized along the center of the nano-wire. Because of these two factors, that is, strong changes in dispersion and localization of the top valance bands, we expect a strong dependence of conductance on the lateral size of the Si nano-wire.

We now return to the overlap between the evanescent  wavefunctions $\psi_{\pm k}({\bm r})$ and $\Delta V_{\mbox{\tiny{L/R}}}$:  
\begin{equation}
\Psi_{\mbox{\tiny{L/R}}}({\bm r}) = \psi_{\mp k}({\bm r})\Delta V_{\mbox{\tiny{L/R}}}({\bm r}).
\end{equation}
Since $\psi_{\mp k}({\bm r})$ decays exponentially inside the leads and $\Delta V_{\mbox{\tiny{L/R}}}$ decays extremely fast inside the alkyl chain, the two overlaps $\Psi_{\mbox{\tiny{L/R}}}({\bm r})$ are exponentially localized at the left/right contacts, respectively. As a consequence the spectral operator in Eq.~\ref{spectral} is only needed in a region near the contacts, which is an important numerical observation. But $\Psi_{\mbox{\tiny{L/R}}}({\bm r})$ also contains important physical information. An iso-surface plot of $\Psi_{\mbox{\tiny{L/R}}}({\bf r})$ is shown in Fig.~\ref{psi}(a), on top of the atomic structure of the device. This plot allows us to understand how different Si atoms contribute to the contact conductance. First thing we see in Fig.~\ref{psi}(a) is that  $\Psi_{\mbox{\tiny{L/R}}}({\bm r})$ is very localized in the lateral direction therefore revealing a clear path inside the Si nano-wires which marks the region which is important to the contact conductance. Fig.~\ref{psi}(a) also gives us a quantitative way to assess how deep into the Si nano-wires the overlap extends. For even better quantitative assessment, we show in Fig.~\ref{psi} the xy average of  $\Psi_{\mbox{\tiny{L/R}}}({\bm r})$. According to this graph, the contact conductance is determined by the first four Si layers of the nano-wires.

\section{Conclusions} 

In conclusion, we have extended the theory of the off-resonance tunneling transport to finite temperatures. The extension enables first-principles calculations of the charge transport in molecular devices connected to semi-conducting leads. We presented an application of the formalism to devices made of alkyl chains connected to Si nano-wires.

The transport characteristics of the devices were characterized  as functions of temperature and number of monomers in the alkyl chains. It was found that, for energy values above the valence states, the transmission decays exponentially with the number of monomers. The exponential decay constant is determined by the imaginary part of the complex wave-vector of the alkyl chains evaluated at the top of the valence band of the Si nano-wires. 
Using an empirical model of the band bending, we compared our results with experimental data and showed that the predicted I-V curves capture all the qualitative features of the experimental curves.
 
The transport characteristics of the devices were further analyzed by mapping the physical quantities entering in our analytic formula for transmission. We found a strong overlap between the evanescent waves of the alkyl chain and the local density of states at the very top of the valence band, leading to an enhanced contribution of these states to the transmission of the devices. Quantitative assessments of the localization of the contact conductance were also provided.

\begin{acknowledgments} 
This research was supported by a Cottrell award from the Research Corporation for Science Advancement and by the office of the Provost of Yeshiva University.
\end{acknowledgments}

%\bibliography{../../TDFT}

\end{document}